\documentclass[%
longbibliography,
preprint,
amsmath,amssymb,
aps,
pra,
]{revtex4-1}
\usepackage{graphicx}
\usepackage{dcolumn}
\usepackage{bm}
\usepackage{epsfig}
\usepackage{afterpage}

\begin{document}

\title{Temperature Dependent Density Profiles of Dipolar Droplets}

\author{E. Aybar}
\email{enes.aybar@bilkent.edu.tr}
\author{M. \"{O}. Oktel}
 \email{oktel@fen.bilkent.edu.tr}
\affiliation{Department of Physics, Bilkent University, Ankara 06800, Turkey}
\date{\today}

\begin{abstract}
Recently, trapped dipolar gases were observed to form high density droplets in a
regime where mean field theory predicts collapse. These droplets
present a novel form of equilibrium where quantum fluctuations are critical for
 stability. So far, the effect of quantum fluctuations have only been considered at zero temperature
 through the local chemical potential arising from the Lee--Huang--Yang correction.
 Here, we extend the theory of dipolar droplets to non-zero
 temperatures using Hartree--Fock--Bogoliubov theory (HFBT), and show that the
 equilibrium is strongly affected by temperature fluctuations.
 HFBT, together with local density approximation for excitations,
 reproduces the zero temperature results, and predict that the condensate density
 can change dramatically even at low temperatures where the total
 depletion is small. Particularly, we find that typical experimental
 temperatures ($T \sim $ 100 nK) can significantly modify the transition between
 low density and droplet phases.
\end{abstract}
\maketitle

\section{Introduction}
\label{section:introduction}

Experiments on ultracold atoms with dipole-dipole interactions provide
opportunities to explore novel physical regimes. So far, Bose-Einstein
condensates where dipolar interaction plays a dominant role have been achieved for
chromium \cite{ChromiumBEC}, dysprosium \cite{DysprosiumBEC} and
erbium \cite{ErbiumBEC}. The long range and anisotropic interaction make
these systems non--trivial and susceptible to catastrophic collapse \cite{PfauReview}. Recent
experiments have surprisingly found that dipolar gases have a stable
droplet phase in a parameter range where mean field theory predicts
collapse \cite{Rosensweig,ErbiumDroplet}.

Formation of stable dipolar droplets were first reported by the Stuttgart group
\cite{Rosensweig}. Subsequent experiments were able to isolate single droplets
\cite{LHYDipolarExperimentalProof}, and show that they can be stable even
without external trapping \cite{PfauSelfBound}. Similarly, the phase transition
between trapped cloud and the droplet has been explored for erbium
\cite{ErbiumDroplet}.

Mean field theory in the form of Gross--Pitaevskii (GP) approximation have been
successfully used to explain the physics of ultracold bosonic systems including
dipolar BECs \cite{ExactHydrodynamics}. However, GP equation predicts collapse
of dipolar BECs in the regime tested by the droplet experiments
\cite{RotonInstability}. Hence, the stability of droplets must either stem from
higher order interactions \cite{Blakie3Body,Saito3Body}, or beyond mean field
effects \cite{LHYMixture,WachtlerSantos}. Experiments have clearly demonstrated
that beyond mean field effects are better candidates for the stability mechanism
\cite{LHYDipolarExperimentalProof}. Quantum fluctuations included as a local
Lee-Huang-Yang (LHY) chemical potential correction \cite{OriginalLHY} has
successfully explained experimentally observed phase transition
\cite{BlakieTrapped}. Although this energy correction is small compared to the
mean field terms, it is crucial for the equilibrium observed in the droplet
phase.

While it is intuitively appealing to include the energy cost of quantum
fluctuations as a local change in the chemical potential, this approach is not
transparent as to which approximations are made in its derivation. There are
systematic approximation methods to calculate the effect of quantum fluctuations
on mean field equations \cite{HFBApproximation}. In this paper, we use HFBT to
take the feedback effect of fluctuations on the condensate into account.
Fluctuations are described by the Bogoliubov--de Gennes (BdG) equations, and we
show that solving BdG equations locally reproduces the generalized GP approach
used in the current literature \cite{WachtlerSantos,BlakieTrapped}. The success
of this equation to explain the experiments is then seen to be a clear
consequence of the depleted density being much lower than the condensate
density. We also show that, contrary to a recent claim \cite{ebucuma},
HFBT approach is enough by itself to describe the droplet phase,
without the ad-hoc inclusion of the LHY term in the chemical potential. 

Generally, the density profile of a BEC depends only weakly on the temperature as long as it
is small compared to the transition temperature \cite{Stringari}. Even the  collective oscillation
frequencies of BECs are modified by temperature only if there is a significant
thermal component in the cloud \cite{ExcitationFreq}. Thus, the density profile of the
condensate is generally calculated within the GP approximation without any
reference to the temperature. In this paper, we show that this is no longer true
for the dipolar clouds close to the droplet transition. When the stability of
the system is provided by fluctuations, temperature effects become non
negligible. HFBT is easily generalized to non zero temperatures, and
clearly shows that the LHY local term can be modified significantly by
temperature even if the total depletion remains small.

This paper is organized as follows: We first discuss the HFBT approach starting
from the Hamiltonian, and then solve BdG equations within the local density
approximation. These approximations yield the generalized GP equation
\cite{WachtlerSantos,BlakieTrapped} up to a small correction. Subsequently, we
discuss the relevant temperature scales in the experiments and calculate how the
LHY term depends on the temperature. Finally, we use this theory to investigate the
dependence of the density profile on temperature and argue that temperature
effects could be relevant in the current experiments.

\section{Hartree-Fock-Bogoliubov Theory}
\label{section:HFBT}

The Hamiltonian for a trapped dipolar Bose gas is:
\begin{eqnarray}\label{Hamiltonian} \hat{H}&=&\int d^{3} \mathbf{x}
\hat{\psi}^{\dagger}(\mathbf{x}) h_{0}(\mathbf{x}) \hat{\psi}(\mathbf{x})\\ & +
&\frac{1}{2} \iint d^{3} \mathbf{x} d^{3} \mathbf{x}'
\hat{\psi}^{\dagger}(\mathbf{x})
\hat{\psi}^{\dagger}(\mathbf{x}')V_{\text{int}}(\mathbf{x}-\mathbf{x}')
\hat{\psi}(\mathbf{x}') \hat{\psi}(\mathbf{x}), \nonumber
\end{eqnarray}
where the Bosonic field operators satisfy
$[\hat{\psi}(\mathbf{x}),\hat{\psi}^{\dagger}(\mathbf{x}')]=\delta(\mathbf{x}-\mathbf{x}')$.
Single particle Hamiltonian
$h_{0}(\mathbf{x})=-\frac{\hbar^{2}\nabla^{2}}{2M}+U_{\text{tr}}(\mathbf{x})-\mu$,
contains the kinetic energy, trapping potential $U_{\text{tr}}(\mathbf{x})$ and
the chemical potential $\mu$. The particles interact through short range
repulsion $g=4\pi\hbar^{2}a_{s}/M$ and long range dipolar potential,
$V_{\text{int}}(\mathbf{x})=g\left[\delta(\mathbf{x})
+\frac{3\epsilon_{dd}}{4\pi
|\mathbf{x}|^{3}}\left(1-3\frac{z^2}{|\mathbf{x}|^{2}}\right) \right]$, where
$\epsilon_{dd}=C_{dd}/3g$ is the dimensionless dipole interaction strength
expressed in terms of s-wave scattering length $a_{s}$.


In the existence of a macroscopically occupied condensate state ($N-N_{0}\ll N$,
where $N$ is the total number of atoms, and $N_{0}$ is the number of condensate
atoms), the field operator can be approximated by a classical mean field plus
fluctuations: $ \hat{\psi}(\textbf{x})=\Psi(\textbf{x})+\hat{\phi}(\textbf{x})
$. These fluctuation operators, $\hat{\phi}$, satisfy the commutation relations,
$\left[\hat{\phi}(\textbf{x}),\hat{\phi}^{\dagger}(\textbf{x}')\right] =
\delta(\textbf{x}-\textbf{x}')-\Psi(\textbf{x})\Psi^{*}(\textbf{x}')/N_{0} $ and
$ \left[\hat{\phi}(\textbf{x}),\hat{\phi}(\textbf{x}')\right] =
\left[\hat{\phi}^{\dagger}(\textbf{x}),\hat{\phi}^{\dagger}(\textbf{x}')\right]=0
$. Then, the non-condensate densities, direct and anomalous, are given by
$\tilde{n}(\textbf{x}',\textbf{x})=\langle\hat{\phi}^{\dagger}(\textbf{x}')\hat{\phi}(\textbf{x})\rangle
$ $(\tilde{n}(\textbf{x})=\langle\hat{\phi}^{\dagger}(\textbf{x})\hat{\phi}(\textbf{x})\rangle)$, and
$\tilde{m}(\textbf{x}',\textbf{x})=\langle\hat{\phi}(\textbf{x}')\hat{\phi}(\textbf{x})\rangle$. As our focus
is the stabilization of the condensate due to fluctuations, we will not
perturbatively expand in the fluctuation operators, but consider their feedback
on the condensate \cite{HFBApproximation}.
Hartree-Fock-Bogoliubov theory includes third and higher order terms via
Hartree-Fock factorization \cite{HFBApproximation}. When applied to third order terms in the
Hamiltonian, this factorization generates:
  \begin{align}
  \hat{\phi}^{\dagger}(\textbf{x})\hat{\phi}^{\dagger}(\textbf{x}')
  \hat{\phi}(\textbf{x}') &\approx
 \tilde{m}^{*}(\textbf{x}',\textbf{x})\hat{\phi}(\textbf{x}') +
\tilde{n}^{*}(\textbf{x}',\textbf{x})\hat{\phi}^{\dagger}(\textbf{x}') +
   \tilde{n}(\textbf{x}')\hat{\phi}^{\dagger}(\textbf{x}) \nonumber \\
    \hat{\phi}^{\dagger}(\textbf{x})\hat{\phi}(\textbf{x}')
  \hat{\phi}(\textbf{x}) &\approx
  \tilde{n}^{*}(\textbf{x}',\textbf{x}) \hat{\phi}(\textbf{x})  +
  \tilde{n}(\textbf{x}) \hat{\phi}(\textbf{x}') +\tilde{m}(\textbf{x}',\textbf{x})  \hat{\phi}^{\dagger}(\textbf{x})  \nonumber \\
    \hat{\phi}^{\dagger}(\textbf{x})\hat{\phi}^{\dagger}(\textbf{x}')
  \hat{\phi}(\textbf{x}) &\approx
 \tilde{m}^{*}(\textbf{x}',\textbf{x})\hat{\phi}(\textbf{x}) +
  \tilde{n}(\textbf{x})\hat{\phi}^{\dagger}(\textbf{x}') +
  \tilde{n}(\textbf{x}',\textbf{x})\hat{\phi}^{\dagger}(\textbf{x}) \nonumber\\
      \hat{\phi}^{\dagger}(\textbf{x}')\hat{\phi}(\textbf{x}')
  \hat{\phi}(\textbf{x}) &\approx
  \tilde{n}(\textbf{x}') \hat{\phi}(\textbf{x})  +
  \tilde{n}(\textbf{x}',\textbf{x}) \hat{\phi}(\textbf{x}')  +
  \tilde{m}(\textbf{x}',\textbf{x})  \hat{\phi}^{\dagger}(\textbf{x}').
\end{align}

The Hamiltonian, then, consists of terms of zeroth, first and second order in
fluctuations. In the many particle ground state the first order terms in fluctuations must vanish.
Therefore, the condensate wavefunction must obey the Gross-Pitaevskii equation:
\begin{equation}
  \mathcal{L}\Psi(\textbf{x})+\int d^{3}\textbf{x}'V_{\text{int}}(\textbf{x}-\textbf{x}')
  \tilde{n}(\textbf{x}',\textbf{x})\Psi(\textbf{x}')+
  \int d^{3}\textbf{x}'V_{\text{int}}(\textbf{x}-\textbf{x}')\tilde{m}(\textbf{x}',\textbf{x})\Psi^{*}(\textbf{x}')=0,
 \end{equation}
 where
$  \mathcal{L}=\left[-\hbar^{2}\nabla^{2}/2M-\mu+U_{\text{tr}}(\textbf{x})+\int d^{3}\textbf{x}'V_{\text{int}}(\textbf{x}-\textbf{x}')|\Psi(\textbf{x}')|^{2}
  +\int
  d^{3}\textbf{x}'V_{\text{int}}(\textbf{x}-\textbf{x}')\tilde{n}(\textbf{x}')\right],$ includes not only
the single particle Hamiltonian, but also the Hartree potential
$\Phi_{H}(\mathbf{x})=\int d^{3}\mathbf{x}'
V_{\text{int}}(\mathbf{x}-\mathbf{x}')(|\Psi(\mathbf{x}')|^{2}+\tilde{n}(\mathbf{x}'))$.
Fluctuation terms generate the direct non-condensate density
$\tilde{n}(\mathbf{x}',\mathbf{x})=\langle\hat{\phi}^{\dagger}(\mathbf{x}')
\hat{\phi}(\mathbf{x})\rangle$,
and the anomalous non-condensate density
$\tilde{m}(\mathbf{x}',\mathbf{x})=\langle\hat{\phi}(\mathbf{x}')\hat{\phi}(\mathbf{x})\rangle$.

Excitation modes and energies are found via the diagonalization of the Hamiltonian.
Although the fourth order terms in the Hamiltonian can be reduced to second
order ones via the Hartree--Fock factorization, we neglect these terms
since they solely involve the interaction among the depleted particles. Such terms
are important only if the depleted density is comparable to the condensate density.
The Hamiltonian is diagonal in the quasiparticle excitations given by the Bogoliubov
transformation:
\begin{align}
 \hat{\phi}(\textbf{x})&=\sum_{j} u_{j}(\textbf{x})\hat{\alpha}_{j}-v^{*}_{j}(\textbf{x})
 \hat{\alpha}^{\dagger}_{j} \nonumber\\
 \hat{\phi}^{\dagger}(\textbf{x})&=\sum_{j} u^{*}_{j}(\textbf{x})\hat{\alpha}^{\dagger}_{j}-v_{j}(\textbf{x})
 \hat{\alpha}_{j},
\end{align}
where, $\hat{\alpha}$ are the quasiparticle operators satisfying
$  \left[\hat{\alpha}_{j},\hat{\alpha}^{\dagger}_{k}\right]=\delta_{j,k}$ and
$ \left[\hat{\alpha}_{j},\hat{\alpha}_{k}\right]=
  \left[\hat{\alpha}^{\dagger}_{j},\hat{\alpha}^{\dagger}_{k}\right]=0$.
  This transformation yields the Bogoliubov-de Gennes equations:
\begin{equation}\label{eq:BdGu}
  \mathcal{L}_{0} u_{j}(\textbf{x})+\int d^{3}\textbf{x}'V_{\text{int}}(\textbf{x}-\textbf{x}')
  \Psi^{*}(\textbf{x}')\Psi(\textbf{x}) u_{j}(\textbf{x}')
 -\int d^{3}\textbf{x}'V_{\text{int}}(\textbf{x}-\textbf{x}')
\Psi(\textbf{x}')\Psi(\textbf{x}) v_{j}(\textbf{x}')
 =E_{j}u_{j}(\textbf{x})
 \end{equation}
 \begin{equation}\label{eq:BdGv}
  \mathcal{L}_{0} v_{j}(\textbf{x})+\int d^{3}\textbf{x}'V_{\text{int}}(\textbf{x}-\textbf{x}')
  \Psi(\textbf{x}')\Psi^{*}(\textbf{x}) v_{j}(\textbf{x}')
 -\int d^{3}\textbf{x}'V_{\text{int}}(\textbf{x}-\textbf{x}')
\Psi^{*}(\textbf{x}')\Psi^{*}(\textbf{x}) u_{j}(\textbf{x}')
 =-E_{j}v_{j}(\textbf{x}),
 \end{equation}
 where $  \mathcal{L}_{0}=\left[-\hbar^{2}\nabla^{2}/2M-\mu+U_{\text{tr}}(\textbf{x})+\int d^{3}\textbf{x}'V_{\text{int}}(\textbf{x}-\textbf{x}')|\Psi(\textbf{x}')|^{2}\right].$
 Bogoliubov amplitudes further satisfy,\\
$ \int d^{3}\textbf{x}\left[u_{j}^{*}(\textbf{x})u_{k}(\textbf{x})-v^{*}_{j}(\textbf{x})v_{k}(\textbf{x})\right]=\delta_{j,k}$, and
$\int d^{3}\textbf{x}\left[u_{j}(\textbf{x})v_{k}(\textbf{x})-u_{k}(\textbf{x})v_{j}(\textbf{x})\right]=0$.

Since the excitation modes are decoupled, the following expectation values are given by Bose statistics,
 \begin{eqnarray}
   \left\langle\hat{\alpha}^{\dagger}_{j} \hat{\alpha}_{k}\right\rangle&=&\delta_{j,k}N_{B}(E_{j}) \nonumber\\
   \left\langle\hat{\alpha}_{j} \hat{\alpha}_{k}\right\rangle&=&\left\langle\hat{\alpha}^{\dagger}_{j}
   \hat{\alpha}^{\dagger}_{k}\right\rangle=0,
 \end{eqnarray}
where
   $N_{B}(E)=1\left/\left(\exp\left[\frac{E}{k_{B}T}\right]-1\right)\right.$. This yields
   temperature dependent depletion density expressions:
 \begin{equation}
   \tilde{n}(\textbf{x}',\textbf{x})=\sum_{j}\left(v_{j}(\textbf{x}')v_{j}^{*}(\textbf{x})
   +N_{B}(E_{j})\left[u_{j}^{*}(\textbf{x}')u_{j}(\textbf{x})+v_{j}(\textbf{x}')v_{j}^{*}(\textbf{x})\right]\right)
 \end{equation}
 \begin{equation}
   \tilde{m}(\textbf{x}',\textbf{x})=-\sum_{j}\left(u_{j}(\textbf{x}')v_{j}^{*}(\textbf{x})
   +N_{B}(E_{j})\left[v_{j}^{*}(\textbf{x}')u_{j}(\textbf{x})+u_{j}(\textbf{x}')v_{j}^{*}(\textbf{x})\right]\right).
 \end{equation}

In principle, a numerical solution of the above set 
would determine both the condensate density and the excitation frequencies.  However, such a determination of
stability is computationally expensive, and numerical approaches so far required further approximations.
For example  in \cite{BohnFinite} the normal density matrix is assumed to be diagonal real space $ \tilde{n}(\textbf{x}',\textbf{x})\propto \delta(\textbf{x}',\textbf{x})$,
which misses most of the dipolar contribution to the local LHY potential. This approximation is repeated in \cite{ebucuma}, and the LHY term is added separately to the BdG equation. 
Simpler approaches based on the generalized GP provide more insight as well as quantitative predictions in line
with the droplet experiments \cite{WachtlerSantos,BlakieTrapped,ErbiumDroplet}. HFB theory introduces three new terms into the GP equation: the direct interaction
between condensed atoms and depleted atoms,
\begin{equation}
  \Phi_{H}^{(1)}(\textbf{x})=\int
  d^{3}\textbf{x}'V_{\text{int}}(\textbf{x}-\textbf{x}')\tilde{n}(\textbf{x}'),
  \end{equation}and the fluctuation terms,
\begin{align}
 \Omega^{(n)}(\textbf{x})\Psi(\textbf{x})&=\int d^{3}\textbf{x}'V_{\text{int}}(\textbf{x}-\textbf{x}')
  \tilde{n}(\textbf{x}',\textbf{x})\Psi(\textbf{x}')\\
  \Omega^{(m)}(\textbf{x})\Psi(\textbf{x})&=\int d^{3}\textbf{x}'V_{\text{int}}(\textbf{x}-\textbf{x}')
  \tilde{m}(\textbf{x}',\textbf{x})\Psi^{*}(\textbf{x}').
\end{align}
These fluctuations can be interpreted as local corrections for the chemical
potential
$\Delta\mu(\textbf{x})=\Omega^{(n)}(\textbf{x})+\Omega^{(m)}(\textbf{x})$.
Therefore, the generalized GP equation becomes:
\begin{equation}\label{eq:GP}
 \Big[-\frac{\hbar^{2}\nabla^{2}}{2M}-\mu+U_{\text{tr}}(\textbf{x})+\Phi_{H}(\textbf{x})
 +\Delta\mu(\textbf{x})\Big]\Psi(\textbf{x})=0,
\end{equation}
where
$\Phi_{H}(\textbf{x})=\int
d^{3}\textbf{x}'V_{\text{int}}(\textbf{x}-\textbf{x}')\left(|\Psi(\textbf{x}')|^{2}+\tilde{n}(\textbf{x}')\right).
$
In the next section we show that the local evaluation of these terms result in the generalized GP equation used in the literature without
any further assumptions. HFBT combined with local density approximation for fluctuations results in the generalized GP equation directly, 
no ad-hoc terms are needed for the description of the stable droplet. 

\section{Local Density Approximation}
\label{section:LDA}

In this section, we give two results which arise when the local density approximation is applied to the HFBT theory given in the previous section.
First, when LDA is applied to BdG equations fluctuation modes can be analytically obtained which reduce the GP equation to the modified GP currently used in the 
literature to describe the droplets.  The second result is that this analysis, including the LDA, can be straightforwardly generalized to non-zero temperatures.

If the condensate density and the trapping potential vary slowly on the scale of
the wavelength of the BdG modes, Eqs.\ref{eq:BdGu},\ref{eq:BdGv} can be solved with a local density
approximation \cite{Pelster} in the spirit of the semi-classical WKB approximation. This approximation gets more accurate for higher
energy modes which makes it more suitable for finite size systems like droplets.

Under the assumption that the condensate density is a slowly
varying function of position, one substitutes \cite{Pelster}
\begin{equation}
u_{j}(\textbf{x}) \rightarrow
u(\textbf{x},\textbf{k})e^{i\textbf{k}\cdot\textbf{x}} \qquad E_{j} \rightarrow
E(\textbf{x},\textbf{k}) \qquad \sum_{j}\rightarrow \int
\frac{d^{3}\textbf{k}}{(2\pi)^{3}}, \end{equation} where
$u(\textbf{x},\textbf{k})$ is also a slowly varying function of position. The
orthogonality condition for the excitation amplitudes then reads
$|u(\textbf{x},\textbf{k})|^{2}-|v(\textbf{x},\textbf{k})|^{2}=1$. The
fluctuation terms can be expressed within the same LDA as \begin{eqnarray}
\Omega_{n}(\textbf{x})= \int \frac{d^{3}\textbf{k}}{(2\pi)^{3}}
\tilde{V}_{\text{int}}(\textbf{k}) \left(|v(\textbf{x},\textbf{k})|^{2}+
N_{B}(E(\textbf{x},\textbf{k}))
\left[|u(\textbf{x},\textbf{k})|^{2}+|v(\textbf{x},\textbf{k})|^{2} \right]
\right)\\ \Omega_{m}(\textbf{x})= \int \frac{d^{3}\textbf{k}}{(2\pi)^{3}}
\tilde{V}_{\text{int}}(\textbf{k})
\left(-u(\textbf{x},\textbf{k})v^{*}(\textbf{x},\textbf{k})-2N_{B}(E(\textbf{x},\textbf{k}))
u(\textbf{x},\textbf{k})v^{*}(\textbf{x},\textbf{k}) \right), \end{eqnarray}
where
$\tilde{V}_{\text{int}}(\textbf{k})=g[1+\epsilon_{dd}(3\cos^{2}\theta_{\textbf{k}}-1)]
$ is the Fourier transform of the interaction potential. Using,
$e^{-i\textbf{k}\cdot\textbf{x}}\mathcal{L}
u(\textbf{x},\textbf{k})e^{i\textbf{k}\cdot\textbf{x}}\approx
\varepsilon_{\textbf{k}} u(\textbf{x},\textbf{k})$, and \begin{eqnarray}
\Psi(\textbf{x})\int d^{3}\textbf{x}'V_{\text{int}}(\textbf{x}-\textbf{x}')
\Psi(\textbf{x}')u(\textbf{x}',\textbf{k})e^{-i\textbf{k}\cdot(\textbf{x}-\textbf{x}')}&=&\Psi(\textbf{x})\int
\frac{d^{3}\textbf{k}'}{(2\pi)^3} \int
d^{3}\textbf{x}'\tilde{V}_{\text{int}}(\textbf{k}')\Psi(\textbf{x}')u(\textbf{x}',\textbf{k})
e^{-i(\textbf{k}-\textbf{k}')\cdot(\textbf{x}-\textbf{x}')}\nonumber\\ &\approx&
n_{0}(\textbf{x})\tilde{V}_{\text{int}}(\textbf{k})u(\textbf{x},\textbf{k}),
\end{eqnarray}

the BdG equations simplify to the algebraic form of:
\begin{equation}
  \varepsilon_{\textbf{k}}u(\textbf{x},\textbf{k})+n_{0}(\textbf{x})\tilde{V}_{\text{int}}(\textbf{k})u(\textbf{x},\textbf{k})-n_{0}(\textbf{x})\tilde{V}_{\text{int}}(\textbf{k})v(\textbf{x},\textbf{k})=
 E(\textbf{x},\textbf{k})u(\textbf{x},\textbf{k})
 \end{equation}
 \begin{equation}
  \varepsilon_{\textbf{k}}v(\textbf{x},\textbf{k})+n_{0}(\textbf{x})\tilde{V}_{\text{int}}(\textbf{k})v(\textbf{x},\textbf{k})-n_{0}(\textbf{x})\tilde{V}_{\text{int}}(\textbf{k})u(\textbf{x},\textbf{k})=
 -E(\textbf{x},\textbf{k})v(\textbf{x},\textbf{k}),
 \end{equation}
 where $\varepsilon_{\textbf{k}}=\frac{\hbar^{2}\textbf{k}^{2}}{2M}$, and
 $n_{0}(\textbf{x})=|\Psi(\textbf{x})|^{2}$.
 Then, the energy spectrum reads:
\begin{equation}\label{eq:spectrum}
   E(\textbf{x},\textbf{k})=\sqrt{\varepsilon_{\textbf{k}}
   \left(\varepsilon_{\textbf{k}}+2n_{0}(\textbf{x})\tilde{V}_{\text{int}}(\textbf{k})\right)}.
 \end{equation}

Thus within the LDA, the modes are labeled by a momentum $\mathbf{k}$ at each position
$\mathbf{x}$ with energy
$E(\mathbf{x},\mathbf{k})=\sqrt{\varepsilon_{\mathbf{k}}
(\varepsilon_{\mathbf{k}}+2n_{0}(\mathbf{x})\tilde{V}_{\text{int}}(\mathbf{k}))}$,
where  $\tilde{V}_{\text{int}}(\mathbf{k})=g[1+\epsilon_{dd}(3\cos^{2}\theta_{k}-1)]$.
Bogoliubov amplitudes are, then, given by \begin{eqnarray}\label{amplitudes}
&|v(\mathbf{x},\mathbf{k})|^{2}=\left(\varepsilon_{\mathbf{k}}+n_{0}(\mathbf{x})
\tilde{V}_{\text{int}}(\mathbf{k})-E(\mathbf{x},\mathbf{k})\right)/2E(\mathbf{x},\mathbf{k})
\nonumber\\
&u(\mathbf{x},\mathbf{k})v^{*}(\mathbf{x},\mathbf{k})=n_{0}(\mathbf{x})
\tilde{V}_{\text{int}}(\mathbf{k})/2E(\mathbf{x},\mathbf{k}). \end{eqnarray} 

Let us first focus on the case of zero temperature. As the 
fluctuation amplitudes are expressed in terms of the local condensate density,
Eq. \ref{eq:GP} becomes a self-consistent equation only for the wavefunction,
\begin{eqnarray}\label{mGP} \left[h_{0}+\Phi_{H}(\mathbf{x})
+\Omega_{n}(\mathbf{x}) +\Omega_{m}(\mathbf{x}) \right] \Psi(\mathbf{x})=0,
\end{eqnarray}
where the usual GP equation is modified by terms caused by
fluctuations. These terms can be evaluated within the same LDA used for the solution
of the BdG equations. With appropriate renormalization \cite{Pelster}
\begin{widetext} \begin{eqnarray}\label{QFterms}
\Omega_{n}(\textbf{x})\Psi(\textbf{x})&\approx& \Psi(\textbf{x})\int
\frac{d^{3}\textbf{k}}{(2\pi)^{3}}\tilde{V}_{\text{int}}(\textbf{k})
|v(\textbf{x},\textbf{k})|^{2}=\frac{8}{3}gn_{0}(\textbf{x})\sqrt{\frac{a^{3}_{s}n_{0}(\textbf{x})}{\pi}}
\mathcal{Q}_{5}(\epsilon_{dd})\Psi(\textbf{x}), \\
\Omega_{m}(\textbf{x})\Psi(\textbf{x})&\approx&-\Psi(\textbf{x})\int
\frac{d^{3}\textbf{k}}{(2\pi)^{3}}\tilde{V}_{\text{int}}(\textbf{k})
u(\textbf{x},\textbf{k})v^{*}(\textbf{x},\textbf{k})=8gn_{0}(\textbf{x})\sqrt{\frac{a^{3}_{s}n_{0}(\textbf{x})}{\pi}}
\mathcal{Q}_{5}(\epsilon_{dd})\Psi(\textbf{x}), \nonumber \end{eqnarray} where $
Q_{l}(\epsilon_{dd})=\int_{0}^{1} du [1+\epsilon_{dd}(3u^{2}-1)]^{l/2} $. As a
result, we obtain the generalized GP equation
\cite{WachtlerSantos,BlakieTrapped}, plus a correction due to the Hartree
potential created by the depleted particles. \begin{equation}\label{explicitGP}
\left[-\frac{\hbar^{2}\nabla^{2}}{2M}+U_{\text{tr}}(\mathbf{x})-\mu+\int
d^{3}\mathbf{x}'V_{\text{int}}(\mathbf{x}-\mathbf{x}')\left(|\Psi(\mathbf{x}')|^{2}+\tilde{n}(\mathbf{x}')\right)+\frac{32}{3}g\sqrt{\frac{a^{3}_{s}}{\pi}}
\mathcal{Q}_{5}(\epsilon_{dd})|\Psi(\mathbf{x})|^{3}\right]\Psi(\mathbf{x})=0.
\end{equation} \end{widetext}
As the depletion $\tilde{n}(\mathbf{x})=\frac{8}{3}\sqrt{\frac{a^{3}_{s}}{\pi}}
\mathcal{Q}_{3}(\epsilon_{dd})|\Psi(\mathbf{x}')|^{3}$ remains small in the droplet
experiments, the extra term in the Hartree potential can be neglected as in the
current literature. It is important to stress that the modified GP equation above is systematically derived 
from HFBT without ad-hoc considerations about the nature of the local chemical potential.

 Still, it is remarkable for two reasons that the LHY local correction, $\Delta
\mu_{QF} (\mathbf{x})=\frac{32}{3}g\sqrt{\frac{a^{3}_{s}}{\pi}}
\mathcal{Q}_{5}(\epsilon_{dd})|\Psi(\mathbf{x})|^{3}$ is exactly reproduced by
the HFBT method. First, contrary to claim in ref.\cite{ebucuma} although HFBT is a mean field theory it can describe a stable droplet phase. While the fluctuations stabilize the
droplet, they are not critical in the renormalization group sense. Any approach that takes the feedback between condensate and fluctuations even at the mean field level
can describe a stable droplet. Second,  
the commonly used Popov approximation neglects the anomalous
density terms  to describe the long wavelength gapless modes correctly
\cite{HFBApproximation}. However, in a finite size system such as the droplets,
the contribution of short wavelength modes are more important, and $3/4$ of the
local LHY chemical potential is provided by the anomalous term. While Popov approximation is commonly employed  in 
numerical calculations of trapped cloud densities\cite{Stringari,ebucuma}, it  underestimates the LHY correction at zero temperature by a factor of 4. Hence
quantitatively accurate description of dipolar droplets cannot be obtained within the Popov approximation.

Apart from giving a systematic derivation of the generalized GP equation, the
HFBT can be generalized straightforwardly to non--zero
temperatures. For the short range interacting trapped Bose condensates,
the effect of temperature on the density profile is negligibly small, and is
mainly caused by interaction with the thermal cloud \cite{Stringari}. However, for the current
droplet experiments, the equilibrium is contingent upon the compressibility
provided by the quantum fluctuations. For a system at finite temperature local fluctuations are provided from both virtual and thermal exctitations.
Temperature fluctuations can compliment quantum fluctuations, and strongly modify the equilibrium. HFBT method
directly identifies how the LHY term in the generalized GP depends on the temperature.

The effect of temperature is easily introduced in terms of the diagonal
operators as $\langle \hat{\alpha}^{\dagger}_{j} \hat{\alpha}_{k} \rangle
=\delta_{j,k} N_{B}(E_{j})$, with $N_{B}(E)=1/(\exp[E/k_{B}T]-1)$. Thus, the
thermal contribution to the LHY correction becomes:
\begin{eqnarray}\label{Gamma_Th}
\Delta \mu_{Th}(\mathbf{x})&=&\int
\frac{d^{3}\mathbf{k}}{(2\pi)^{3}}\tilde{V}_{\text{int}}(\mathbf{k})
N_{B}(E(\mathbf{x},\mathbf{k}))\\ &\times &\left[
|v(\mathbf{x},\mathbf{k})|^{2}+
|u(\mathbf{x},\mathbf{k})|^{2}-
2u(\mathbf{x},\mathbf{k})v^{*}(\mathbf{x},\mathbf{k})\right].\nonumber
\end{eqnarray}

It is instructive to identify two different temperature scales for an
interacting BEC. For a weakly interacting system at zero temperature, the number
of the atoms in the condensate is much larger than the number of depleted atoms.
As the temperature is increased, more atoms leave the condensate. The total
number of depleted atoms is comparable to the number of atoms in the condensate
if the temperature is near the BEC critical temperature. However, at a much
lower temperature, the number of thermally depleted atoms will be comparable to
the number of depleted atoms at zero temperature. If the presence of the depleted
atoms is a determining factor for the equilibrium state as in the droplet
experiments, temperature will start to affect the condensate density at these
lower temperatures. Thus, temperature effects can be important even if the total
depleted density is small compared to the condensate.

\begin{figure} \includegraphics[width=0.7\textwidth]{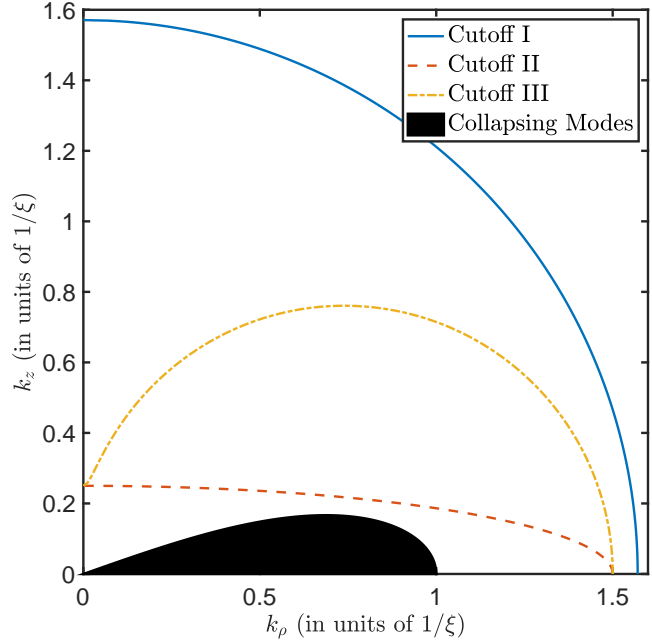}
\caption{Cutoff I, is the cutoff used in this paper which has an isotropic from of
$k_{c}=\pi/2\xi$.
Cutoff II is the cutoff used in \cite{BlakieTrapped} which is given by
$k_{c}(\vartheta)=1/\sqrt{\sin^{2}\vartheta/k_{c,\rho}^{2}+\cos^{2}\vartheta/k_{c,z}^{2}}$.
Cutoff III is the cutoff used in \cite{WachtlerSantos} which is given by
$k_{c}(\vartheta)=\sqrt{k_{c,\rho}^{2} \sin^{2}\vartheta
+k_{c,z}^{2}\cos^{2}\vartheta}$, $\{k_{c,\rho},k_{c,z}\}=\{1.5,0.25\}\xi^{-1}$ for both options.
Blackened region is the modes with imaginary energies when $\epsilon_{dd}=1.5$.
} \label{fig:cutoff} \end{figure}

For an infinite homogenous system, if the dipolar interaction is dominant
$(\epsilon_{dd}>1)$, the quasi particle energy becomes imaginary in a region of
$\mathbf{k}$-space, signaling an instability. If the local density approximation is
strictly applied to the LHY correction, an imaginary term will appear
in the generalized GP equation. However, these unstable modes are long
wavelength in character and they are the principal cause of the formation of the
droplet state. Thus, for a finite size droplet, the wavelength of these modes
are least the size of the system. The finite size effect can be incorporated
into the LDA by choosing a cutoff in $\mathbf{k}$-space. Different choices of
cutoff parameters were seen to give small changes in the LHY correction as most of the
contribution comes from short wavelength modes \cite{BlakieTrapped,WachtlerSantos}.
 Hence, we consider a spherical cutoff in $\mathbf{k}$-space with inverse coherence length of the
condensate $k_{c}=\frac{\pi}{2\xi}$. This choice is physically motivated for LDA by $\xi$ being
the length scale over which the condensate density is essentially constant.In the literature, one finds two other cutoff choices: Ref. \cite{BlakieTrapped} uses
 an elliptical cutoff, $k_{c}^{(II)}(\vartheta)=1/\sqrt{\sin^{2}\vartheta/k_{c,\rho}^{2}+\cos^{2}\vartheta/k_{c,z}^{2}}$; and Ref. \cite{WachtlerSantos} uses the cutoff, $k_{c}^{(III)}(\vartheta)=\sqrt{k_{c,\rho}^{2} \sin^{2}\vartheta
 +k_{c,z}^{2}\cos^{2}\vartheta}$. Moreover, in the energy spectrum given by Eq. \ref{eq:spectrum}, the density of
 states at zero energy is finite for $\epsilon_{dd}>1$. The existence of a cutoff is more crucial for non-zero temperature calculations
because the density of states at zero energy becomes finite for
$(\epsilon_{dd}>1)$. Using the cutoff to exclude only the unstable modes would
result in a logarithmic divergence in thermal fluctuations. In Fig. \ref{fig:cutoff}, we plot these cutoff choices
 as well as the region of imaginary modes in the k-space. We see that (Fig. 1,
 in text) all of these cutoff choices yield similar results.

\begin{figure} \includegraphics[width=0.7\textwidth]{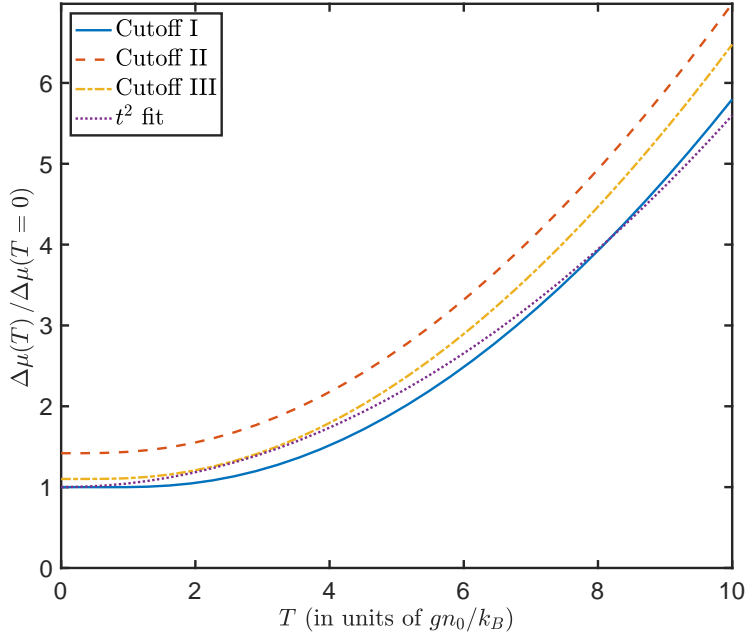} \caption{
Temperature dependence of local LHY correction on the unitless temperature,
 $t =k_{B}T/gn_{0}$, calculated with different cutoff options for $\epsilon_{dd}=1.5$.
Cutoff I is the spherical cutoff employed in this paper $\left(k_{c}^{\text{(I)}}=\pi/2\xi\right)$
(blue dotted line), Cutoff II, $k_{c}(\vartheta)^{\text{(II)}}=1/\sqrt{\sin^{2}\vartheta/k_{c,\rho}^{2}
+\cos^{2}\vartheta/k_{c,z}^{2}}$ (orange dashed line), and Cutoff III,
$k_{c}(\vartheta)^{\text{(III)}}=\sqrt{k_{c,\rho}^{2} \sin^{2}\vartheta +k_{c,z}^{2}\cos^{2}\vartheta}$
(yellow dash-dotted line), where $\{k_{c,\rho},k_{c,z}\}=\{1.5,0.25\}\xi^{-1}$are the anisotropic
 cutoffs used in \cite{BlakieTrapped} and \cite{WachtlerSantos} respectively. The $t^{2}$ fit used
 in the energy functional (Eq. \ref{eq:Functional}) for the Cutoff I is also plotted (purple solid line).}
\label{fig:Q5}
\end{figure}

 Hence, at finite temperature the Bogoliubov amplitudes:
 \begin{eqnarray}\label{eq:BdGamplitudes}
 |v(\mathbf{x},\mathbf{k})|^{2}&=&\frac{\varepsilon_{\textbf{k}}+n_{0}(\mathbf{x})\tilde{V}_{\text{int}}(\mathbf{k})}{2E(\mathbf{x},\mathbf{k})}-\frac{1}{2}\nonumber\\
 u(\mathbf{x},\mathbf{k})v^{*}(\mathbf{x},\mathbf{k})&=&\frac{n_{0}(\mathbf{x})\tilde{V}_{\text{int}}(\mathbf{k})}{2E(\mathbf{x},\mathbf{k})}
 \end{eqnarray}
 give the  correction terms:
 \begin{equation}
\Omega_{n}(\textbf{x})= \int \frac{d^{3}\textbf{k}}{(2\pi)^{3}} \tilde{V}_{\text{int}}(\textbf{k})
\Big\{\frac{\varepsilon_{\mathbf{k}}+n_{0}(\mathbf{x})\tilde{V}_{\text{int}}(\mathbf{k})
-E(\mathbf{x},\mathbf{k})}{2E(\mathbf{x},\mathbf{k})}
+N(E(\textbf{x},\textbf{k})) \frac{\varepsilon_{\mathbf{k}}+n_{0}(\mathbf{x})\tilde{V}_{\text{int}}(\mathbf{k})}{E(\mathbf{x},\mathbf{k})} \Big\}
\end{equation}
\begin{equation}
\Omega_{m}(\textbf{x}) = \int \frac{d^{3}\textbf{k}}{(2\pi)^{3}} \tilde{V}_{\text{int}}(\textbf{k})
\Big\{-\frac{n_{0}(\mathbf{x})\tilde{V}_{\text{int}}(\mathbf{k})}{2E(\mathbf{x},\mathbf{k})}
+\frac{n_{0}(\mathbf{x})\tilde{V}_{\text{int}}(\mathbf{k})}{2\varepsilon_{\mathbf{k}}}
-N(E(\textbf{x},\textbf{k})) \frac{n_{0}(\mathbf{x})\tilde{V}_{\text{int}}(\mathbf{k})}{E(\mathbf{x},\mathbf{k})} \Big\},
\end{equation}
where the second term is properly renormalized. The local LHY correction becomes
\begin{equation}
\Delta\mu(\textbf{x}) = \int \frac{d^{3}\textbf{k}}{(2\pi)^{3}} \tilde{V}_{\text{int}}(\textbf{k})
 \left\{ \frac{\varepsilon_{\mathbf{k}}}{2E(\mathbf{x},\mathbf{k})}
 +\frac{n_{0}(\mathbf{x})\tilde{V}_{\text{int}}(\mathbf{k})}{2\varepsilon_{\mathbf{k}}}-\frac{1}{2} +\frac{1}{\exp\left[E(\mathbf{x},\mathbf{k})/k_{B}T\right]-1}\frac{\varepsilon_{\mathbf{k}}}{E(\mathbf{x},\mathbf{k})} \right\}.
 \end{equation}
Using \begin{equation}
\xi(\textbf{x})=\sqrt{\frac{\hbar^{2}}{2Mgn_{0}(\textbf{x})}},
\end{equation}
$k=q/\xi$, $\cos\vartheta=u$, $f(u)=1+\epsilon_{dd}\left(3u^{2}-1\right)$, and
$t(\textbf{x})=\frac{k_{B}T}{gn_{0}(\textbf{x})}$, one can write
\begin{align}
  \Delta\mu(\textbf{x}) = \frac{g}{4\pi^{2} \xi^{3}(\textbf{x})} \int_{-1}^{1}du \int_{q_{c}}^{\infty}q^{2}dq f(u)
  \Big\{& \frac{q^{2}}{2\sqrt{q^{2}\left(q^{2}+2f(u)\right)}}+\frac{f(u)}{2q^{2}}-\frac{1}{2} \\
  &+\frac{1}{\exp\left[\sqrt{q^{2}\left(q^{2}+2f(u)\right)}/t(\textbf{x})\right]-1} \frac{q^{2}}{\sqrt{q^{2}\left(q^{2}+2f(u)\right)}}
  \Big\}.\nonumber
  \end{align}
  Since $\xi\propto\Psi^{-1}$, the local change in the chemical potential is
  \begin{equation}
\Delta\mu(\mathbf{x}) = \frac{32}{3}g\sqrt{\frac{a_s^{3}}{\pi}}\left(\mathcal{Q}_{5}(\epsilon_{dd})
+\mathcal{R}(\epsilon_{dd},t(\mathbf{x}))\right)|\Psi(\textbf{x})|^{3} .
\end{equation}
Unitless functions $\mathcal{Q}_{5}$ and $\mathcal{R}$ are given by
\begin{equation} \mathcal{Q}_{5}(\epsilon_{dd};q_{c})=\frac{1}{4\sqrt{2}}\int_{0}^{1}du f(u)\left[\left(4f(u)-q_{c}^{2}\right)\sqrt{2f(u)+q_{c}^{2}}-3f(u)q_{c}+q_{c}^{3}\right] \end{equation}
\begin{equation} \mathcal{R}(\epsilon_{dd},t;q_{c})=\frac{3}{4\sqrt{2}}\int_{0}^{1}du\int_{q_{c}^{2}}^{\infty}dQ\frac{Qf(u)}{\sqrt{Q+2f(u)}}\frac{1}{\exp[\sqrt{Q\left(Q+2f(u)\right)}/t]-1}. \end{equation}

\begin{figure} \includegraphics[width=0.7\textwidth]{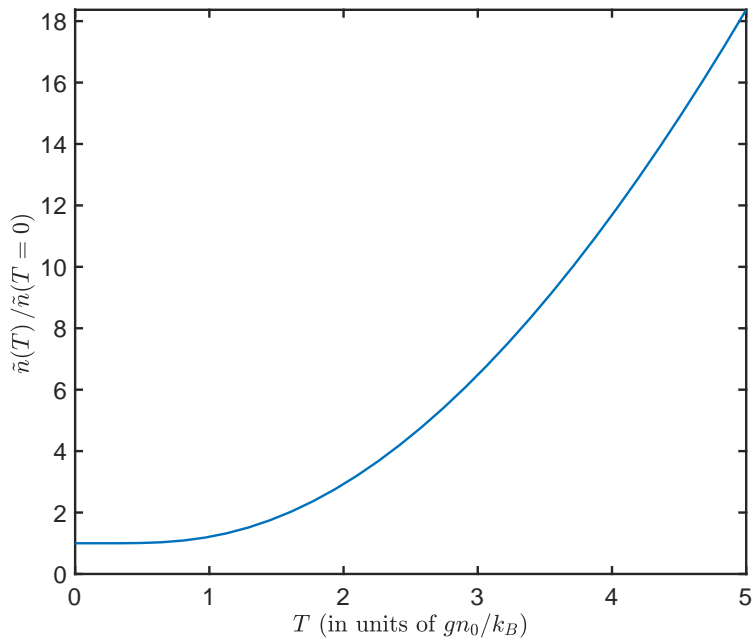}
\caption{Non-condensate density $\tilde{n}(\mathbf{x})$ as a function of unitless temperature, $t=k_{B}T/g n_0$.} \label{fig:depletion}
\end{figure}

Within the same LDA, the depleted density is given by
\begin{equation}
\tilde{n}(\textbf{x})= \int \frac{d^{3}\textbf{k}}{(2\pi)^{3}}
\left(|v(\textbf{x},\textbf{k})|^{2}+ N_{B}(E(\textbf{x},\textbf{k}))
\left[|u(\textbf{x},\textbf{k})|^{2}+|v(\textbf{x},\textbf{k})|^{2} \right]
\right). \end{equation} Using the Bogoliubov amplitudes given in Eq.
\ref{eq:BdGamplitudes}, one finds \begin{equation} \tilde{n}(\mathbf{x}) =
\frac{8}{3}g\sqrt{\frac{a_s^{3}}{\pi}}\left(\mathcal{Q}_{3}(\epsilon_{dd})
+\mathcal{P}(\epsilon_{dd},t(\mathbf{x}))\right)|\Psi(\textbf{x})|^{3}, \end{equation}
where \begin{equation}
\mathcal{Q}_{3}(\epsilon_{dd};q_{c})=\frac{1}{\sqrt{2}}\int_{0}^{1}du
f(u)\left[\left(f(u)-q_{c}^{2}\right)\sqrt{2f(u)+q_{c}^{2}}+q_{c}^{3}\right]
\end{equation} \begin{equation}
\mathcal{P}(\epsilon_{dd},t;q_{c})=\frac{3}{\sqrt{2}}\int_{0}^{1}du\int_{q_{c}^{2}}^{\infty}dQ\frac{Q+f(u)}{\sqrt{Q+2f(u)}}\frac{1}{\exp[\sqrt{Q\left(Q+2f(u)\right)}/t]-1}.
\end{equation} The non-condensate density increases with increasing temperature due
to thermal depletion. In Fig. \ref{fig:depletion}, temperature dependence of the
non-condensate density is plotted. It is important to note that, near the edge of the condensate
 the unitless temperature increases as the condensate density
decreases. Although the fraction of the
non-condensate to the condensate density increases near the edge, total number of depleted atoms can remain small.

In the regime where the non-condensate density is negligible compared to condensate
density, the generalized GP becomes:
\begin{eqnarray}
  \Big[h_{0}+\int d^{3} \mathbf{x}' V_{\text{int}}(\mathbf{x}-\mathbf{x}')|\Psi(\mathbf{x}')|^{2}
     +  \Delta \mu(\mathbf{x})
  \Big]\Psi(\mathbf{x})=0,
\end{eqnarray}
where $\Delta \mu(\mathbf{x})$ encompasses both quantum and thermal fluctuations:
\begin{equation}
\Delta \mu(\mathbf{x}) = \frac{32}{3}g\sqrt{\frac{a_s^{3}}{\pi}}\left(\mathcal{Q}_{5}(\epsilon_{dd})
+\mathcal{R}(\epsilon_{dd},t)\right)|\Psi(\mathbf{x})|^{3}.
\end{equation}
Temperature fluctuation term $\mathcal{R}$ depends on the unitless temperature $t=k_{B}T/gn_{0}$.
In Fig. \ref{fig:Q5}, we display the temperature dependence of LHY correction for our cutoff
choice. We check that other cutoff choices yield similar temperature
dependencies. 

In the next section we concentrate on the solution of this modified GP equation, particularly highlighting the effect of
dramatic consequences of small but non-zero temperatures.

\section{Variational Calculation of Temperature Dependent Density Profiles} 
\label{section:Variational}

As a first step to estimate the effects of temperature dependent LHY correction,
we employ a Gaussian variational ansatz. 
Energy functional corresponding to the generalized GP
equation (Eq. \ref{eq:GP}) is similar to what is used in Ref. \cite{BlakieTrapped}. However, the
thermal fluctuation term, $\mathcal{R}$, depends on condensate density through the
unitless temperature. To get an analytical form for energy functional in $\Psi$,
we used a power low fit for the $\mathcal{R}$ function. A $t^{n}$ curve for $n>2.5$
results in a divergence near the condensate edge where the condensate density is
low and the unitless temperature is high. This divergence, however, is a
byproduct of the Gaussian variational method, where the condensate extends to
infinity. We find that a $t^{2}$ fit describes numerically obtained values within $0<t<10$ and results in a finite correction even when integrated
over all space. In Fig. \ref{fig:Q5}, we plot this fit with the function
$\mathcal{R}$. The fit parameter in $\mathcal{R}(\epsilon_{dd},t)=S(\epsilon_{dd})t^{2}$ is found to be 
$\mathcal{S}(\epsilon_{dd})=-0.01029\epsilon_{dd}^{4}+0.02963\epsilon_{dd}^{3}-
0.05422\epsilon_{dd}^{2}+0.009302\epsilon_{dd}+0.1698$ for $0<\epsilon_{dd}<2$.

Therefore, in the region where the depleted density is negligible compared to the
condensate density, the generalized GP equation reads:
\begin{align}
  \left[-\frac{\hbar^{2}}{2M}\nabla^{2}+U_{\text{tr}}(\mathbf{x})+\int d^{3} \mathbf{x}' V_{\text{int}}(\mathbf{x}-\mathbf{x}')|\Psi(\mathbf{x}')|^{2}
   +\gamma |\Psi(\mathbf{x})|^{3} +\theta T^{2} \frac{1}{|\Psi(\mathbf{x})|}
  \right]\Psi(\mathbf{x})=\mu\Psi(\mathbf{x}),
\end{align}
where $\gamma =
\frac{32}{3}g\sqrt{\frac{a_s^{3}}{\pi}}\mathcal{Q}_{5}(\epsilon_{dd})$, and
$ \theta = \frac{32}{3}g\sqrt{\frac{a_s^{3}}{\pi}}\frac{k_{B}^{2}}{g^{2}}\mathcal{S}(\epsilon_{dd})
$, and $\mathcal{S}$ is the found from the $t^{2}$ fit.

The energy functional corresponding to the generalized GP equation above is:
\begin{align}
  E[ \Psi ]=& \int  d^{3}\mathbf{x}  \Psi^{*}(\mathbf{x})\left[ -\frac{\hbar^{2}}{2M}\nabla^{2}+U_{\text{tr}}(\mathbf{x}) \right] \Psi(\mathbf{x} )
  \nonumber\\
  &+\frac{1}{2} \int d^{3}\mathbf{x}\int d^{3}\mathbf{x'}  |\Psi(\mathbf{x})|^{2}V_{\text{int}}(\mathbf{x}-\mathbf{x}')|\Psi(\mathbf{x}')|^{2}
  \nonumber\\
  &+\frac{2}{5} \int d^{3}\mathbf{x} \gamma |\Psi(\mathbf{x})|^{5}\nonumber\\
  &+ 2 \int d^{3}\mathbf{x} \theta T^{2} |\Psi(\mathbf{x})|.
\end{align}

To estimate the temperature effects on the condensate density profile, we used
the Gaussian ansatz
\begin{equation}
  \Psi(\mathbf{x}) =
  \sqrt{\frac{8N}{\pi^{3/2}\sigma_{\rho}^{2}\sigma_{z}}}\exp\left[-2\left(\frac{\rho^{2}}{\sigma_{\rho}^{2}}+\frac{z^{2}}{\sigma_{z}^{2}}\right)\right].
\end{equation}
For the trap potential
$U_{\text{tr}}(\mathbf{x})=\frac{1}{2}M\left(\omega_{\rho}^{2}x^{2}+\omega_{\rho}^{2}y^{2}+\omega_{z}^{2}z^{2}\right)$,
energy per particle for the above functional gives
\begin{eqnarray}\label{eq:Functional}
 \frac{E[\sigma_{\rho},\sigma_{z}]}{N}&=&\frac{\hbar^{2}}{M}\left(\frac{2}{\sigma_{\rho}^{2}}+\frac{1}{\sigma_{z}^{2}}\right)+M\left(\frac{\omega_{\rho}^{2}\sigma_{\rho}^{2}}{8}+\frac{\omega_{z}^{2}\sigma_{z}^{2}}{16}\right)\nonumber \\
 &+&\frac{\sqrt{2}}{\pi^{3/2}}g \frac{N}{\sigma_{\rho}^{2}\sigma_{z}} \left[1-\epsilon_{dd}\mathcal{F}(\sigma_{\rho}/\sigma_{z})\right] \nonumber\\
 &+&\frac{2^{12}}{75\sqrt{5}\pi^{11/4}} g \sqrt{a_s^{3}} \left(\frac{N}{\sigma_{\rho}^{2}\sigma_{z}}\right)^{3/2}\mathcal{Q}_{5}\nonumber\\
 &+& \frac{64\pi^{1/4}}{3}\frac{k_{B}^{2}T^{2}}{g}\sqrt{a_s^{3}}\sqrt{\frac{\sigma_{\rho}^{2}\sigma_{z}}{N}}\mathcal{S},
\end{eqnarray}
where
\begin{equation}
  \mathcal{F}(x)=\frac{1+2x^{2}}{1-x^{2}}-\frac{3x^{2}\tanh^{-1}\sqrt{1-x^{2}}}{\left(1-x^{2}\right)^{3/2}}.
\end{equation}

\begin{figure}
\includegraphics[width=0.7\textwidth]{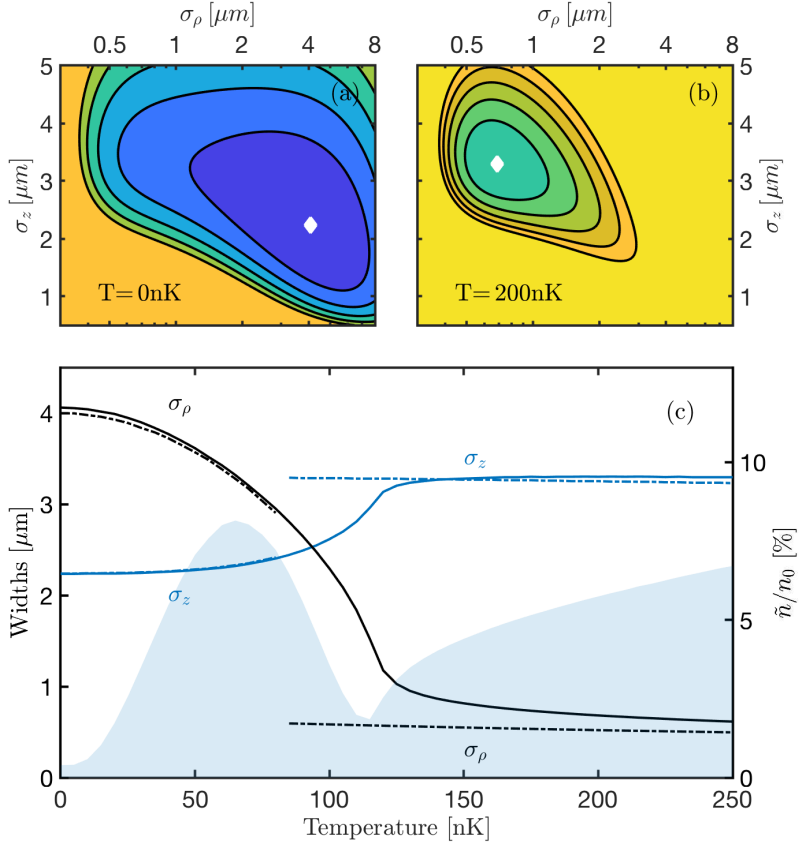} \caption{ (a, b)
Contour plots of total energy calculated with the energy functional Eq.
\ref{eq:Functional} for $2000 ~^{164}$Dy atoms with $a_{dd}=132a_{0}$ and
$a_{s}=93a_{0}$, where $a_{0}$ is the Bohr radius, at temperatures $T =0$ nK and
$T =200$ nK, respectively. White diamonds show the energy minimum for the
Gaussian ansatz. Results are for atoms in a harmonic trap with
$\{\omega_{\rho},\omega_{z}\}=2\pi\times\{45,133\} s^{-1}$. (c) Variational radii of the stable condensate solutions for
$a_{s}=88a_{0}$ (dash-dotted lines) and $a_{s}=93a_{0}$ (solid lines) at
different temperatures for the same parameters as in (a,b).
Shaded area corresponds to the depletion fraction at the center of the condensate calculated for
the $a_{s}=93a_{0}$ case.}
\label{fig:width}
\end{figure}

%
We numerically find $\{\sigma_{\rho},\sigma_{z}\}$ which minimize this energy
functional. Just as the zero temperature case \cite{SantosGroundState,BlakieTrapped} two different
kinds of minima can be observed corresponding to the trapped (low density) and
the droplet (high density) phases. Increasing temperature may cause the system to
shift from trapped phase to the droplet phase. In Fig. \ref{fig:width}, we plot
the radii of the condensate as a function of temperature, for a typical droplet
reported in \cite{Rosensweig}. It is important to note that the transition between the two
phases happens close to $100 $nK, and the total depletion at the center remains
less than $8\%$ throughout.

Stability of self bound droplets \cite{PfauSelfBound} without a trapping potential is solely due to
fluctuations. Hence, thermal fluctuations as well as quantum fluctuations
determine their structure. Temperature dependence of their stability  can be investigated with the same Gaussian
ansatz. To estimate the central density, one writes the chemical potential at
the condensate center
\begin{equation*}
  \left. \mu \right|_{\mathbf{r}=0}= gn_{0}\left(1-\epsilon_{dd}\mathcal{F}(\sigma_{\rho}/\sigma_{z})\right)
  + \gamma n_{0}^{3/2}+\theta T^{2} n_{0}^{-1/2}
\end{equation*}
as in Ref. \cite{LHYDipolarExperimentalProof}. Therefore,
\begin{equation*}
  \left.\frac{\partial\mu}{\partial n_0}\right|_{\mathbf{r}=0}=g\left(1-\epsilon_{dd}\mathcal{F}(\sigma_{\rho}/\sigma_{z})\right)
  +\frac{3}{2}\gamma n_{0}^{1/2}-\frac{1}{2}\theta T^{2} n_{0}^{-3/2}.
\end{equation*}
The stability condition, $\partial\mu/\partial n_0 \geq 0$, yields the equation
for the minimum central density
\begin{equation*}
  0=\alpha+\frac{3}{2}\gamma n_{0}^{1/2}-\frac{1}{2}\theta T^{2} n_{0}^{-3/2},
\end{equation*}
where $\alpha =
g\left(1-\epsilon_{dd}\mathcal{F}(\sigma_{\rho}/\sigma_{z})\right)$. At low
temperatures, treating the temperature term as a perturbation, one gets
\begin{equation}
  \sqrt{n_{0}}=-\frac{2\alpha}{3\gamma}-\frac{9\theta\gamma^{2}}{8\alpha^{3}}T^{2},
\end{equation}
which, then, takes the form
\begin{equation}
  n_{0}(T)=n_{0}(T=0)+\frac{2\mathcal{S}}{3\mathcal{Q}_{5}}\frac{k_{B}^{2}T^{2}}{g^{2}n_{0}(T=0)},
  \end{equation}
  where
   $n_{0}(T=0)=\frac{\pi}{a_{s}^{3}}\left(\frac{\epsilon_{dd}\mathcal{F}(\sigma_{\rho}/\sigma_{z})-1}{16\mathcal{Q}_{5}}\right)^{2}$.

\begin{figure} \includegraphics[width=0.7\textwidth]{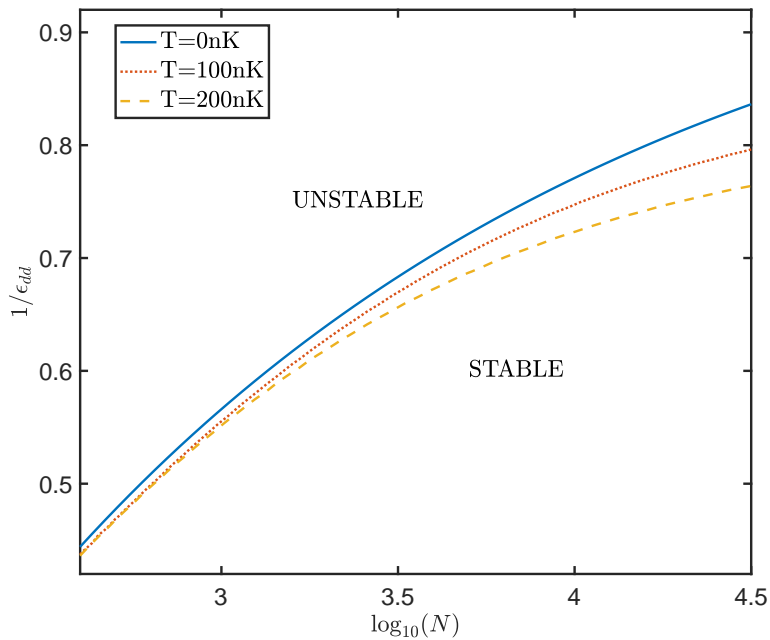}
\caption{Phase diagram for self-bound droplets as a function of $1/\epsilon_{dd}$ and $N$ at $T=0$ nK (blue
solid line), $T=100$ nK (orange dotted line) and $T=200$ nK (yellow dashed
line).}
\label{fig:selfbound}
\end{figure}

In Fig. \ref{fig:selfbound}, we plot the stable
region in particle number and dipolar strength for self bound droplets at
different temperatures. The minimum number of particles required to form a
stable droplet  increases with increasing temperature.

\section{Conclusion}
\label{section:Conclusion}

Let us summarize the main points of the calculation presented in the previous sections and their consequences.
First, we derived the modified GP equation used in the literature to describe the dipolar droplets using  HFBT and LDA applied to fluctuations. This derivation clarifies the assumptions inherent in the modified GP equation, and presents opportunities for systematic improvement. A consequence of this approach is that it constrains successful theoretical descriptions of systems where fluctuations are needed for equilibrium, in particular:
\begin{itemize}
	\item Mean field description, as long as it takes the feedback of fluctuations back on the condensate as in HFBT, can be used to describe such fluctuation stabilized equilibria. 
	\item HFBT equations, solved self consistently for the condensate and fluctuations can describe a stable droplet, without the introduction of ad-hoc terms to the local chemical potential.
	\item Popov approximation, which neglects anomalous non-condensed density is commonly used for trapped gases at finite temperature. However the terms neglected in this approximation provide a significant portion of the feedback on the condensate. Thus quantitatively accurate description of dipolar droplets are not possible within the Popov approximation.
	\item As the dipolar interaction is not short ranged, the correlations in the non-condensed density $\tilde{n}(\mathbf{x}',\mathbf{x})$ are important. Setting $\tilde{n}(\mathbf{x}',\mathbf{x})$ to a delta function before the local density approximation, as is commonly done for finite temperature numerical calculations, is bound to yield quantitatively incorrect results.
\end{itemize}

As a second point, using HFBT equations at finite temperature we generalized the description of dipolar droplets to finite temperatures. Our approach is limited to low enough temperatures so that the number of non-condensed particles are much smaller than the number of particles in the condensate, still our calculations indicate that:
\begin{itemize}
		\item As the novel property of dipolar droplets is their stabilization by fluctuations, they become susceptible to temperature fluctuations even at low temperatures. The temperature scale at which the condensate sufficiently differs from zero temperature is set by comparing the thermally excited particle density with virtually excited particle density, not the condensed density.
		\item Temperature as low as to give a few percent of thermally excited density can drive the transition between trapped and dipolar phases in the current Dy experiments. 
		\item Temperature does not have a straightforward effect on the droplet. While higher temperatures favor increasing density, such as
	the droplet phase over the low density phase in a trapö the minimum number of particles needed to stabilize a droplet also increases with increasing temperature. 
\end{itemize}

Finally, we should outline the limitations of the theory given in this paper and how they can be overcome in future studies. 
First, the use of a variational wavefunction gives a rough measure of stability, but is not expected to be quantitatively correct, particularly in the droplet phase where the density may deviate significantly from a Gaussian. Instead of a variational wavefunction, direct numerical solution of the modified GP equation, including temperature corrections would be more accurate. We will report the results of such simulations in a follow up\cite{enesaybarfurkanozturk}. A second limitation of our calculation is that we neglected the interaction among the non-condensate particles. These interactions can be taken into account by self-consistent numerical solution of BdG equations, still within the LDA. Finally, our use of LDA forces a momentum space cutoff to exclude the unstable solutions. Any approach which takes the discrete nature of BdG modes at low energies into account would remove the need for such an arbitrary cutoff parameter. With such a precise
characterization of temperature dependence, the density profile of dipolar
droplets can be used to probe temperature in the nano-Kelvin regime.

This project is supported by T\"{u}rkiye Bilimsel ve Teknolojik Ara\c{s}t{\i}rma Kurumu
(T\"{U}B\.{I}TAK) Grant No. 116F215.

\bibliography{DipolarDropletsPRA}
\end{document}